\documentclass[letterpaper, twocolumn,superscriptaddress, showpacs,preprintnumbers,amsmath,amssymb] {revtex4}
\usepackage{graphicx}
\usepackage{dcolumn}
\usepackage{bm}

\begin{document}

\title{Interfering directed paths and the sign phase transition}

\author{Hyungwon Kim}
\affiliation{Physics Department, Princeton University, Princeton, NJ 08544, USA}

\author{David A. Huse}
\affiliation{Physics Department, Princeton University, Princeton, NJ 08544, USA}

\begin{abstract}
We revisit the question of the ``sign phase transition'' for
interfering directed paths with real amplitudes in a random medium.  The sign of the
total amplitude of the paths to a given point may be viewed as an Ising order parameter, so we
suggest that a coarse-grained theory for system is a dynamic Ising model
coupled to a Kardar-Parisi-Zhang (KPZ) model.  It appears that when the
KPZ model is in its strong-coupling (``pinned'') phase, the Ising model
does not have a stable ferromagnetic phase, so there is
no sign phase transition.  We investigate this
numerically for the case of $1+1$ dimensions,
demonstrating the instability of the Ising ordered phase there.
\end{abstract}


\maketitle

The problem of interfering directed paths in a random medium was investigated by a few groups
mostly during the 1990's \cite{mksw,gelfand,rc,ng,sff,am}.  It arises in at least two contexts, namely quantum
hopping conduction, and in the long-distance behavior of the high-temperature spin-spin
correlations in a random-exchange spin model (see, e. g. \cite{mksw}).

The model is perhaps simplest to present in the case where
the paths are on a $d+1$ dimensional lattice.  Let $t$ be the coordinate along which the path is directed
(it is convenient to think of it as a time, but it is really just one of the spatial directions).  Each path
then is specified by ${\bf x}(t)$, where this gives the position $\bf x$ of the path in the $d$
transverse dimensions at ``time'' $t$.  The weight $W$ of a single path is the product of the weights $w$
of each of its links:
\begin{equation}
W\{{\bf x}(t)\}=\Pi_t w({\bf x}(t-1),{\bf x}(t);t)~.
\end{equation}
The weights $W$ and $w$ are amplitudes in the case of adding quantum paths, while for the
spin-spin correlations the $w$ are the nearest-neighbor correlations and $W$ is the contribution
to the long-distance correlation due to that path.  We are interested in the case were the
link weights $w$ can be either positive or negative real numbers.
The total weight $Z({\bf x},t)$ of all paths ``arriving'' at
$\bf x$ at ``time'' $t$ is simply generated by the iterative formula
\begin{equation}
Z({\bf x},t)=\Sigma_{{\bf x}'} Z({\bf x}',t-1)w({\bf x}',{\bf x};t)~.
\end{equation}

In the case where all paths have a non-negative real weight $W$, this model is
the statistical mechanics of directed paths in a random potential and $Z$ is then
the partition function of all paths arriving at that point.  This case, with no
negative weights, is well-studied and understood, see, e. g. \cite{fh}.
In this case, if we
define $h({\bf x},t)=-\log{(Z({\bf x},t))}$, then $h$, which is essentially a free
energy, obeys the KPZ equation \cite{kpz} when coarse-grained:
\begin{equation}
\frac{\partial h}{\partial t}=\nu\nabla^2h-\lambda(\nabla h)^2+\eta_h({\bf x},t)~,
\end{equation}
where $\eta_h$ is ``white'' noise that has only short-range correlations in space and time.
The long-distance, long-time behavior of this system has two phases: In $d>2$ there is
a phase where the nonlinearity $\lambda$ is irrelevant, so the dynamics is diffusive.  In
this phase the partition function for the paths is not dominated by a few high-weight paths
and the behavior is asymptotically the same as the
nonrandom case where all paths have equal weight.  For $d\leq 2$ and also for strong enough
randomness in $d>2$, there is a ``pinned'' phase, where $\lambda$ is relevant and the
partition function for long paths is dominated by a few high-weight paths.

When the link weights $w$ have random sign, the behavior is apparently still in the
universality class of the KPZ equation \cite{mksw,gelfand}, at least in the pinned phase.
Presumably the unpinned (or diffusive) phase for $d>2$ can also survive random signs, but it does
not appear that this has been investigated yet.  To consider the case of random signs, we
break the weights $Z$ into sign $s=\pm 1$ and amplitude as $Z({\bf x},t)=s({\bf x},t)\exp{(-h({\bf x},t))}$.
Now consider the case where most of the link weights $w$ are positive, and only a small
fraction of them are negative.  Then the signs of the weights $Z$ at nearby points ${\bf x}$
will tend to be the same, and we can still coarse grain both $h$ and the sign $s$, which
serves as an Ising order parameter here.  The occasional negative link weight $w$ can
flip the sign $s$, thus serving as a noise term for the dynamics of an effective
Ising model.  When this Ising model has a domain wall between positive and negative sign
domains, there is destructive interference at the domain wall, which will increase $h$.
And the difference in $h$ across the domain wall will determine whether the positive or
negative weights dominate in this destructive interference, which causes the domain wall
to move towards the domain with the lower $h$.  As a result of all this, the coarse-grained system becomes
a dynamic Ising model coupled to a KPZ equation as
\begin{equation}
\frac{\partial s}{\partial t}=\nu_s\nabla^2s+rs-us^3-\mu\nabla s\cdot\nabla h+\eta_s({\bf x},t)
\end{equation}
and
\begin{equation}
\frac{\partial h}{\partial t}=\nu_h\nabla^2h-\lambda(\nabla h)^2-\kappa s^2+\eta_h({\bf x},t)~,
\end{equation}
where all the coupling constants shown are positive, and the noise terms are still white.

In the absence of random signs, the sign of $Z$ is uniform in space and time, corresponding
to the ferromagnetic ordered phase of this effective Ising model.  Since there are no spin flips
at all, this corresponds to zero temperature for the Ising model.  An interesting question is
whether this ordered phase survives to a nonzero concentration of negative signs of the $w$'s,
with a ``sign phase transition'' when this long-range space-time order of the sign of $Z$ is lost.
Such a sign phase transition has been demonstrated to occur on certain hierarchical lattices
\cite{am}, and argued to occur even for $d=1$ \cite{sff}, although the latter claim is
doubted by other authors \cite{mksw,rc,ng}.  If this sign transition did indeed occur in
$d=1$, that would be striking, since it would amount to a phase transition at nonzero
temperature in a one-dimensional Ising model.  The usual arguments that would rule this
out do not apply here, since the Ising model is coupled to a nonequilibrium
and nontrivial driven system, namely a KPZ equation.

Thus in this paper we look carefully
at what is happening in this system at a very low concentration of negative values of $w$,
which corresponds to a low nonzero temperature of the corresponding one-dimensional Ising model.
Our result is that there is no sign phase transition for $d=1$, and the reason for this result
seems like it should apply throughout the pinned phase of the KPZ system in any $d$, in
agreement with the conclusion in Ref. \cite{mksw}.  A sign phase transition should be possible
for $d>2$ in the unpinned phase of the KPZ equation, but we leave the investigation of this
for others.

Let's now consider a general one-dimensional dynamic Ising model with a very low rate per unit length $F$ of
spin flips.
When a flip occurs within a very large ``up'' domain, it produces a small ``down'' domain.  The
probability that this new down domain will survive for time $t$ or longer decays with time as
$\sim t^{-p}$ with some power $p$.  For the usual Ising model coupled only to
a heat bath $p=1/2$, but for a driven model like
we are considering, this exponent $p$ can be different.  In fact, it appears that for this problem
of interfering directed paths $p=2/3$, as we will argue and show is consistent with our simulations.
If the domain does survive to time $t$ it will be of typical size $\sim t^{\zeta}$.
For the usual Ising model $\zeta=1/2=p$, while for our system we have the KPZ value
$\zeta=2/3$, and again it appears that $\zeta=p$.  Now consider what happens to a single
infinite up domain after time $t$: the total number of spin flips per unit length that have
happened is $Ft$.  The density of these that produced down domains that survived a time of order
$t$ and thus grew to size of order $t^{\zeta}$ is $\sim Ft^{1-p}$, so the fraction of the line
occupied by these down domains is $\sim Ft^{1+\zeta-p}$. In other words, $\int^{t}_{0}F\tau^{\zeta-p}d\tau \sim Ft^{1+\zeta-p}$. Thus we find that the initial large domain,
and thus the ferromagnetism, is unstable to breaking up in to domains at long time for any nonzero
$F$ provided that $p<1+\zeta$.  For the usual undriven Ising model at a low temperature, this is
indeed true, since $p=\zeta=1/2$, so the ferromagnetic phase is only present in the absence of
spin flips at zero temperature, as is well known.  Note that this instability of the ferromagnetic phase
is not due to typical or median domains, which have a size and lifetime of order one.  It is due to
power-law rare domains with very large lifetime.  Thus in Ref. \cite{sff} when they argue for the presence
of a sign phase transition based on the behavior of typical domains, they are missing the effect
of these rare large domains that destabilize the ferromagnetic phase at any density of negative-weight
links.

For the problem of directed paths in a random potential, we can look first at the limit of
strong pinning where the total weight $Z$ is always dominated by the one path with the largest
weight.  Then if one negative link weight $w$ occurs, the probability that it is on the highest weight
path after time $t$ is given by the amount that the highest weight path wanders.  The highest-weight
path in that vicinity wanders by distance $\sim t^{\zeta}$, so the probability that it passes through the
link with the negative weight, and as a result has a negative weight and makes a surviving down domain of
our effective Ising model decays with exponent $p=\zeta$.  But this is an argument that works in the
limit where only one path dominates in $Z$.  What about for weaker pinning, where many paths can
contribute significantly to $Z$?  To test that regime, we have performed some numerical simulations
of this process.

\begin{figure}
\includegraphics[width=3.50in]{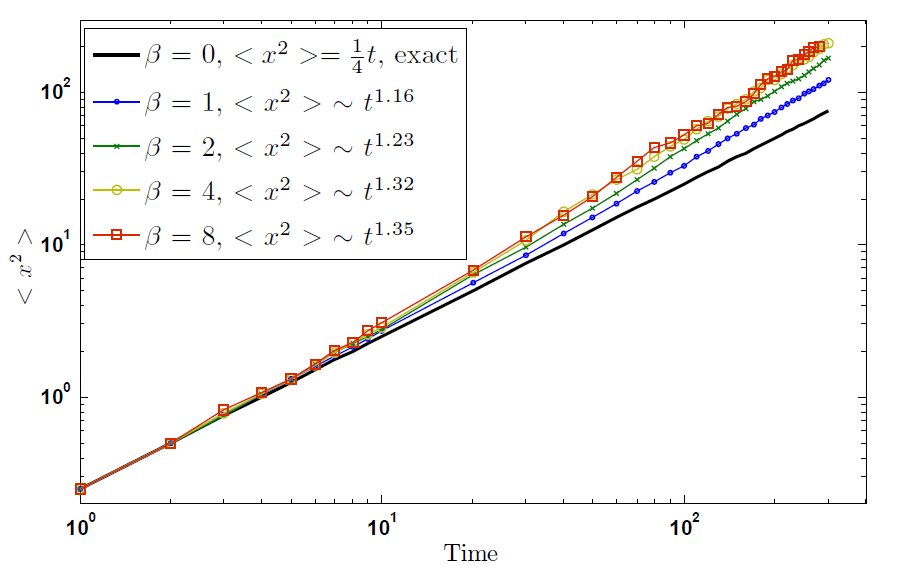}
\centering
\caption{(color online) Mean-square displacement of the paths $\langle x^{2}\rangle$ for various temperatures $\beta=1/T$. The effective exponents in the legend are fit to the ``time'' range 100 to 300.  We want to use a temperature that is intermediate between the $T=0$ behavior at high $\beta$ and the unpinned $\beta=0$ behavior, so we choose to examine $\beta=1$ and $\beta=2$.}
\end{figure}

The model we choose to study has the paths moving on a grid $(x,t)$, with the allowed values of $x$ being any integer
at even-valued times $t$, and $x+\frac{1}{2}$ being any integer at odd-valued times.  The links change $x$ by $\pm\frac{1}{2}$,
so the nonzero values of the link weights are only $w(x,x\pm\frac{1}{2};t)$.  We introduce a temperature $T=1/\beta$;
the link weights are $w(x,x';t)=\pm (u(x,x';t))^{\beta}$, where the $u$'s are
uniformly distributed random numbers between $0$ and $1$.  When we restrict the link weights to be all positive, this
is a directed path in a random potential.  If we start with the path passing through the origin, so
$Z(0, 0) = 1$ and $Z(x \neq 0, 0) = 0$, then for infinite $T$ ($\beta=0$) the paths wander as $\langle x^2(t)\rangle=t/4$, while
at zero $T$ (infinite $\beta$) the lowest energy path wanders as $\langle x^2(t)\rangle\sim t^{4/3}$ at long ``time''.
For the present study, we want to examine intermediate temperatures, where the random pinning is present, but
multiple paths do contribute to $Z$.  In Fig. 1 we show the wandering of the paths vs. $t$ for various $\beta$
for short times $t\leq 300$.  We choose to study $\beta=1$ and $\beta=2$, as these do show intermediate
behavior in this plot at this accessible time range.

\begin{figure}
\includegraphics[width=3.50in]{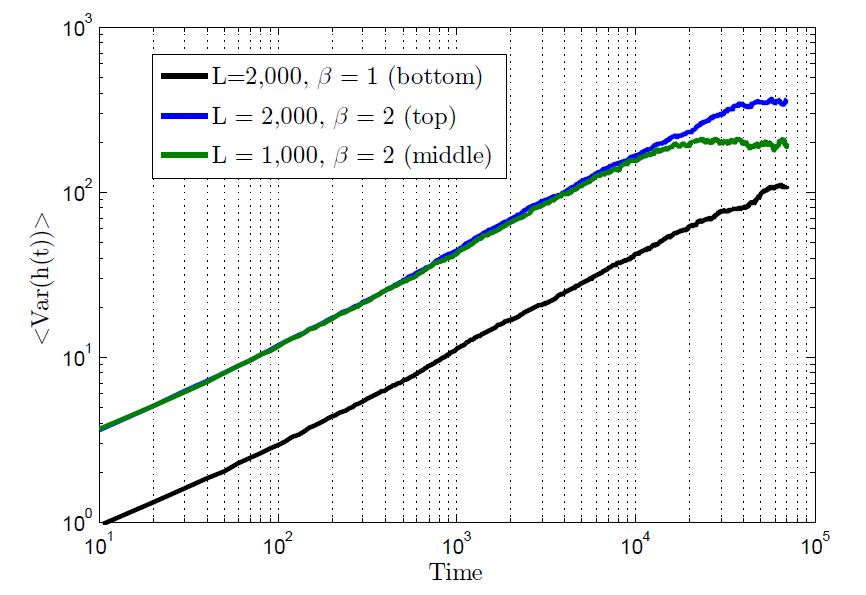}
\centering
\caption{(color online) Variance of the free energy for lattice size $L = 1000$ and 2000 for $\beta = 2$ and $L = 2000$ for $\beta = 1$.
The behavior of the growth of variance in the intermediate time range is independent of $L$.
We can clearly see the saturation due to the finite size effect around $t \sim 20000$ for $L = 1000$ and $\beta=2$.
We averaged over 200 realizations.}
\end{figure}

We will look at the statistics of the dynamics of one ``down'' domain in a large ``up'' domain,
produced by flipping the sign of $Z(x,t)$ at one or a few consecutive site(s) $x$,
while keeping $Z(x',t)$ positive at that time at all other sites, and keeping all the subsequent link weights $w$ positive.
We chose a lattice length $L$ and impose periodic boundary conditions in the ``spatial'' ($x$) direction.
We are interested in the limit of infinite $L$, but for any finite $L$ there will be finite-size effects
when the paths wander around the system and ``notice'' the periodic boundary conditions.
To observe the time scale at which finite-size effects enter, we started with $Z(x,0)=1$ at all sites $x$ and then measured the site-to-site variance of the
free energy, $h(x,t)=-\log{Z(x,t)}$, at each time, $Var(h(t))$, for the case were the link weights are all positive.  The results are shown in Fig. 2. The expected behavior of straight line is seen in an intermediate time range between an early-time transient
and the late time when the finite size effect enters, causing the variance to saturate.  The results for times before the finite-size
effects start are independent of $L$ and thus representative of the infinite system that we want to study.
Since we will be studying just one flipped domain in our system, we want $L$ to be as small as possible
to conserve computer time, but we want $L$ large enough so that we avoid finite size effects.
For the scale of simulations we chose to do, the compromise that
we chose was to study L = 2,000 for both $\beta$ = $1$ and $2$.

\begin{figure}
\includegraphics[width=3.50in]{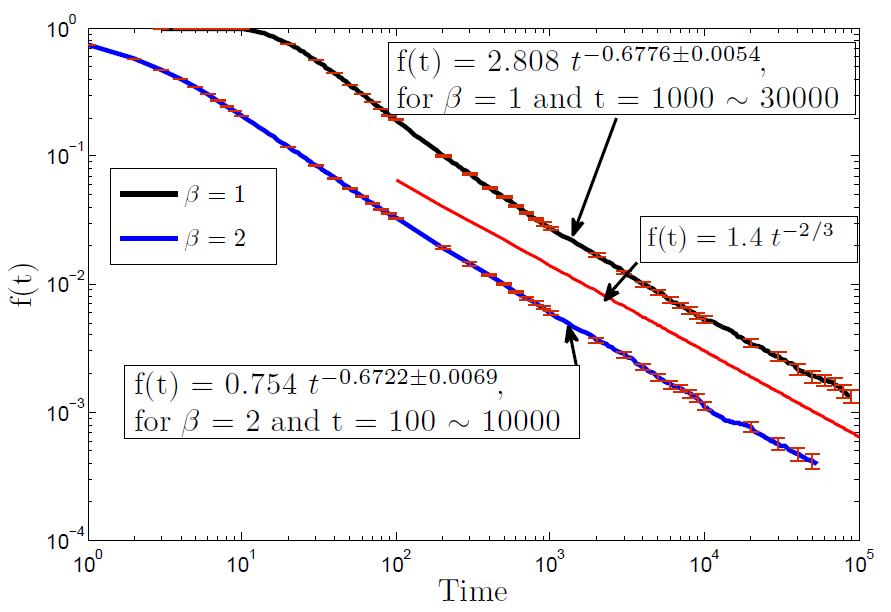}
\centering
\caption{(color online) The survival probability $f(t)$ for $\beta=1$ and 2.
We can clearly see the power law behavior, $f(t) \sim t^{-p}$ with $p = 2/3$, after the initial transient.
A straight line with slope 2/3 is shown for comparison (red line). The ranges for fitting parameters are given within 95 $\%$ confidence level.  }
\end{figure}

We produce and study the domains of negative $Z$ one at a time, thus effectively studying the limit of low
concentration of negative link weights $w$.  In practice we do not introduce any negative link weights,
instead we just flip one or more of the $Z$'s sign(s) ``by hand'' to start a negative domain.
We then follow each such negative domain until it vanishes, or until
it lives so long that finite-size effects enter.  If the domain remains small and vanishes then all the
$Z$'s return to being positive.  If the domain grows to size L, or survives so long we have to ``kill'' it
to not waste computer time, then we change the signs of all the $Z$'s back to positive, without changing
their magnitudes.
Once the negative domain disappears or is removed, we let the system evolve for L/4 time steps to reduce any effects due to
the previous domain.
The vast majority of the domains remain small and vanish before growing to size $L$.  In these cases we start the next domain
at a randomly chosen location that is at a distance at least $L/4$ from the final location of the previous domain, again to
reduce any residual effects of one domain on the next.  We measure the probability $f(t)$ that a domain survives to time $t$,
and the average size (number of ``flipped spins'') $\langle x(t)\rangle$ of the domains that do survive to time $t$.
Due to the relatively greater interference when $\beta = 1$, the survival probability is substantially smaller than for $\beta = 2$,
which is detrimental to the statistics for the large $t$ regime that we are interested in.
Therefore, in order to increase the long-time survival probability,
to initiate a new domain at $\beta=1$ we flipped the sign of $Z$ on four adjacent sites instead of just one,
thus starting with a domain of size four at $t=0$.  For $\beta=2$ the domains are started with size one spin flip.
Since we are interested in the long-time behavior of the domains and large domains dominate at long times, this
small change in the initial conditions does not affect our conclusions, it only changes the number of domains that
survive in to the time range we are interested in.

\begin{figure}
\includegraphics[width=3.50in]{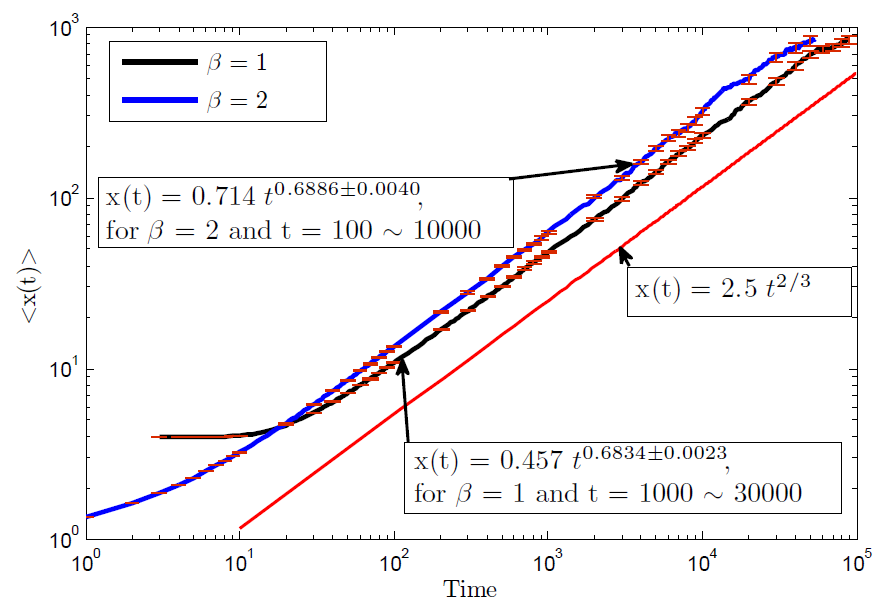}
\centering
\caption{(color online) Expectation value, $\langle x(t)\rangle$, of the size of the negative domains that survived to time $t$.
The straight line has slope 2/3, for comparison (red line). The ranges for fitting parameters are given within 95 $\%$ confidence level.  }
\end{figure}

Our results are shown in Fig. 3 and Fig. 4.
$f(t)$ is simply $N(t)/N(0)$, where $N(0)$ is the number of domains started and $N(t)$ is the number that survived until time $t$.
For $\beta = 2$, $N(0)=143845$; and for $\beta = 1$, $N(0)=54083$.  The small error bars and the clear power-law behavior suggest that we can
conclude that $p \cong \zeta \cong 2/3$ or at least $\textit{very}$ close to 2/3 with confidence.  Thus the long-time scaling behavior
appears to be independent of $\beta$ as long as $\beta >0$.  When $\beta$ is reduced, what happens is that there is a longer
short-time transient as the system crosses over from weak-pinning (low $\beta$) behavior at short times to strong-pinning ($\beta\rightarrow\infty$)
behavior at long times.  We can conclude with confidence that $p < 1+\zeta$ and thus that there is no sign phase transition in $d=1$.

It seems reasonable to expect that a similar behavior will be found in higher $d$ whenever the
system is in the strong-pinning KPZ phase.  There the $d$-dimensional ``volume'' occupied by a
surviving domain grows with exponent $d\zeta$, so the condition for instability of the ferromagnetic phase
at low spin-flip rate, and thus the absence of a sign phase transition is $p < 1+d\zeta$.  But it is exactly this
same volume that sets the survival exponent $p$ when $Z$ can be dominated by one path.  Thus in the strong-pinning KPZ
phase in higher $d$, we expect $p=d\zeta$ and no sign phase transition.

It is notable that $p = \zeta$ for $d=1$ for both the usual Ising model, where $\zeta=1/2$, and for our system of interfering paths with $\zeta=2/3$.
This implies that at long times the average size of all domains, including the domains that have vanished, becomes time-independent.
This is occurring because the two domain walls, once well-separated, are moving independently with no bias to increase or decrease the
size of the domain.  Thus although the finite negative domain has appeared within an infinite positive domain, in the event that the
negative domain grows large, this initial condition is apparently asymptotically forgotten locally at the two domain walls.

We thank Boris Spivak for many discussions and for bringing this
topic to our attention.  We also thank Ben Schaffer for preliminary work, and Kuk Jang for fruitful discussions about numerics.
This work was partially supported by the NSF
through MRSEC grant DMR-0819860 (D.A.H.) and by a Samsung Scholarship (H.K.).

\end{document}